\documentclass[journal=jpclcd,manuscript=letter,layout=onecolumn]{achemso}
\pdfoutput=1
\usepackage{pdfpages}
\usepackage[version=3]{mhchem}          
\usepackage{natmove}
\usepackage{bm}                         
\author{Caterina Cocchi}
\email{caterina.cocchi@unimore.it}
\affiliation{Centro S3, CNR-Istituto Nanoscienze, I-41125 Modena, Italy}
\alsoaffiliation{Dipartimento di Fisica, Universit\`a di Modena e Reggio Emilia, I-41125 Modena, Italy}
\author{Alice Ruini}
\affiliation{Centro S3, CNR-Istituto Nanoscienze, I-41125 Modena, Italy}
\alsoaffiliation{Dipartimento di Fisica, Universit\`a di Modena e Reggio Emilia, I-41125 Modena, Italy}
\author{Deborah Prezzi}
\affiliation{Centro S3, CNR-Istituto Nanoscienze, I-41125 Modena, Italy}
\author{Marilia J. Caldas}
\affiliation{Instituto de F{\'\i}sica, Universidade de S\~ao Paulo, 05508-900 S\~ao Paulo, SP, Brazil}
\author{Elisa Molinari}
\affiliation{Centro S3, CNR-Istituto Nanoscienze, I-41125 Modena, Italy}
\alsoaffiliation{Dipartimento di Fisica, Universit\`a di Modena e Reggio Emilia, I-41125 Modena, Italy}
\title{Designing all-graphene nanojunctions by covalent functionalization}
\begin{document}
\begin{abstract}
We investigated theoretically the effect of covalent edge
functionalization, with organic functional groups, on the electronic
properties of graphene nanostructures and nano-junctions. Our
analysis shows that functionalization can be designed to tune
electron affinities  and ionization potentials of graphene flakes,
and to control the energy alignment of frontier orbitals in
nanometer-wide graphene junctions. The stability of the proposed
mechanism is discussed with respect to the functional groups, their number as well as the width of graphene
nanostructures. The results of our work indicate that different
level alignments can be obtained and engineered in order to realize
stable all-graphene nanodevices.
\end{abstract}
\section*{TOC Graphic}
\begin{figure}%
\centering
\includegraphics[width=9.0cm]{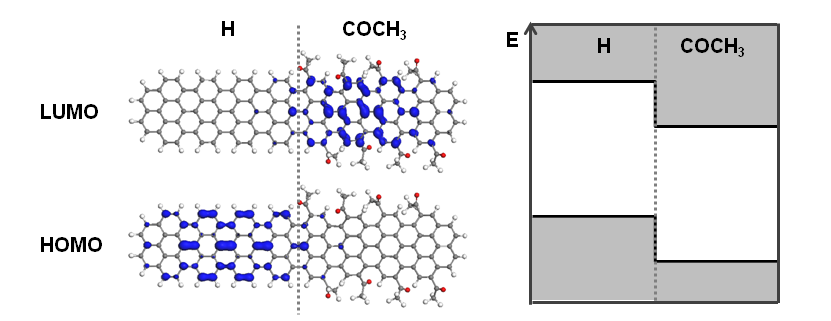}
\end{figure}

\textbf{Keywords}: semi-empirical methods, carbon nanostructures,
frontier orbitals line up, organic moieties

\newpage
The successful fabrication of the two-dimensional system graphene
\cite{novo+04sci} and its excellent mechanical and electronic
properties have aroused great interest in several different fields,
both from fundamental research and for technological applications
\cite{sold+10car,geim09sci,burg+09am,cast+09rmp}.
Quasi-one-dimensional graphene nanoribbons (GNRs) with
semiconducting behavior can also be obtained, \cite{han+07prl,
chen+07pe} with electrical band gap tunable within a wide range,
\cite{son+06prl,yang+07prl} and their optical properties, even if
less intensively studied, offer stimulating perspectives
\cite{prez+08prb,prez+07pssb,prez+10tobe,yang+07nl}. One main
challenge is to gain control of the graphene edge morphology at the
nanoscale, in view of designing all-graphene nano-devices. In this
sense many recent improvements in production techniques were
reported, including both top-down \cite{datt+08nl,
camp+09nl,li+08sci,wang-dai10natc,jia+09sci,kosy+09nat,jiao+09nat,
cano+09nl} and bottom-up \cite{yang+08jacs,cai+10nat} approaches,
which point to the possibility of exploiting the exceptional
properties of graphene nanostructures in actual devices (see e.g.
Refs. \cite{wang+08prl,xia+09natn,pono+08sci}).

In view of electronic and optoelectronic applications, it is
important to identify further approaches for an efficient
engineering of the electronic states  of a graphene nanostructure.
It has recently been shown that the electronic properties of
graphene-based systems can be conveniently modulated by means of
organic functionalization
\cite{loh+10jmc,farm+09nl,farm+09apl,lohm+09nl}. Until now, a number
of procedures have been proposed exploiting either surface
functionalization through chemisorption of molecules
\cite{lope+09nl,beky+09jacs,sini+10nano}, which is expected however
to damage the transport properties of the system, or non-covalent
interaction \cite{chen+07jacs,lu+09jpcb,zhan+10nt, sun+10prb}, which
usually induces doping effects but does not lead to stable
structures. A more effective strategy could be edge
functionalization \cite{qian+08jacs,wang+09sci}, which may offer
promising perspectives also in so far unexplored application fields,
such as the production of GNR-based nanojunctions.

In semiconductor physics, heterojunctions are usually classified on
the basis of the relative alignment of valence and conduction band
states at the interface of the two materials. Heterostructures where
the band gap of one semiconductor is fully contained in the gap of
the other are commonly indicated as \textit{type I} or
\textit{straddling}. On the other hand, the so called \textit{type
II} or \textit{staggered} alignment is achieved when the top of
valence band and the bottom of conduction band of one semiconductor
are both higher in energy compared to the corresponding states of
the other, so that the top of the valence band of the junction lies
on one side of the interface, and the bottom of conduction on the
other side. In this case, electrons and holes on the frontier
orbitals are spatially separated, being located on opposite sides of
the junction, which may promote relevant processes, such as e.g.
photoinduced charge separation, as requested by light harvesting
applications, or photocatalysis. 
These concepts can be extended to graphene nanojunctions, where the 
seamless covalent transition from two graphene nanostructures with 
different electronic characteristics can create, as in conventional 
semiconductor heterojunctions, different kind of alignments. 
Indeed, interfacing semiconducting GNRs with different widths, and 
consequently different energy gaps, provides a \textit{type I} level 
line-up\cite{prez+10tobe,sevi+08prb}. In this paper we
propose a proof of concept for a viable method to realize
all-graphene \textit{type II} nanojunctions. For this purpose we
investigate the effects of stable edge functionalization through
organic electron-donating and electron-withdrawing groups, which
modify electron affinities and ionization potentials in comparison
with the H-terminated GNR.
The proposed model for chemical engineering carries a high potential
for producing stable graphene nanostructures, preserving their
flexibility and their intrinsic electronic features, $sp^2$ carbon network being unaltered upon functionalization.

%
\section{Methods}
\begin{figure}%
\centering
\includegraphics[width=.48\textwidth]{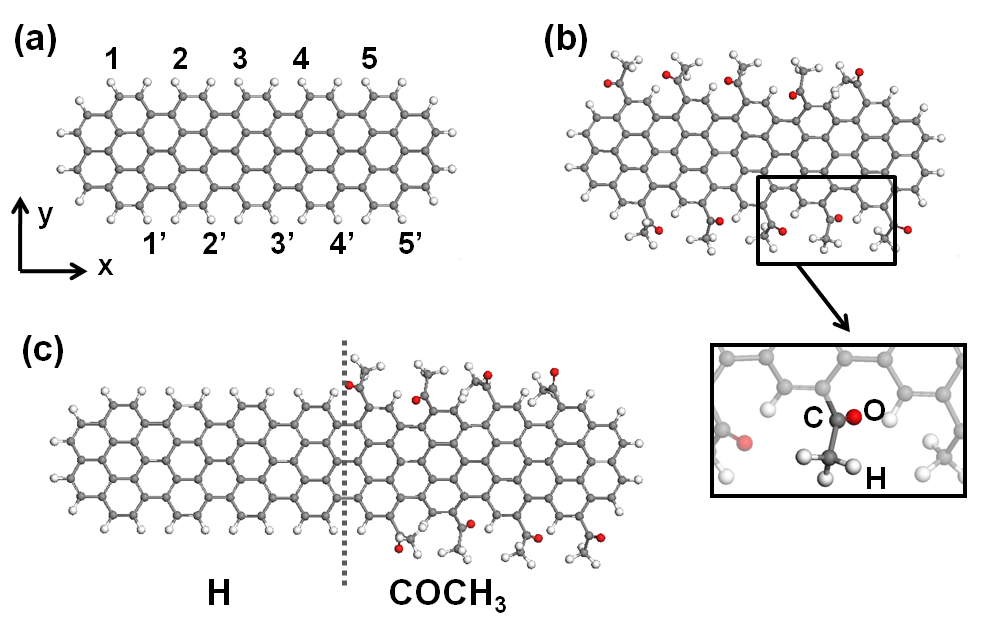}%
\caption{(Color online) (a) Hydrogenated \ce{C80H28} graphene
nanoflake (GNF); the numbers identify the hydrogen atoms that can be
substituted by the organic groups upon functionalization. (b) Fully
functionalized GNF obtained by replacing each second H atom
terminating the edge [i.e. 1-5 and 1'-5' in panel (a)] with a
\mbox{-\ce{COCH3}} radical; in the inset the atoms of the radical
are highlighted.
(c) Nanojunction obtained by fully functionalizing one side of a \ce{C122H40} GNF with \mbox{-\ce{COCH3}} groups.}%
\label{fig1}
\end{figure}

As prototypes for graphene nanoribbons we consider
sub-nanometer-wide graphene nanoflakes (GNFs), that is, finite
systems, and we focus in particular on H-terminated GNFs
[\ref{fig1}(a)], organically functionalized GNFs [\ref{fig1}(b)], as
well as on the nanojunction obtained by functionalizing one half of
the flake with -\ce{COCH3} groups [\ref{fig1}(c)]. In this case, we
will be interested in the frontier molecular orbitals (that stand
here for the conduction and valence band states) and their
localization and energy alignment across the differently terminated
regions. In order to have access to the electronic energies, we
performed calculations in the frame of the semi-empirical AM1 scheme
\cite{dewa+85jacs,AM1-note}. AM1 is a well tested method based on
the Hartree-Fock scheme \cite{root51rmp}, suitable for finite
structures;
\cite{cald+01apl,davi-cald02jcc,lair+08car,tach08jpcc,hant+09ms,abe+10jjap,wetm+00cpl}
it allows one to evaluate the ground state properties of the system
and to compute the electrical band gap $E_{G}$ as the difference
between electron affinity (EA) and ionization potential (IP). The
latter quantities are expressed in terms of the total energy of the
neutral [E(0)] and charged systems [E($\pm$1)], in the usual way EA
= E(0) - E(-1) and IP = E(+1) - E(0). For all the considered
structures, we performed full geometrical optimization to evaluate
E(0), with a threshold for the forces of 0.4 $\text{kcal} \cdot
\text{mol}^{-1}$/\AA{}. 
The E($\pm$1) are obtained at the fixed
atomic positions optimized in the ground state, i.e.
\textit{vertical} EA and IP are computed.
%
\section{Results and Discussion}
Our reference system is the \ce{C80H28} flake [\ref{fig1}(a)],
characterized by armchair-shaped H-saturated edges along the length
(\textit{x} direction in \ref{fig1}). We checked that the chosen
length/width ratio is such that the zigzag-shaped end-borders
(\textit{y} direction) do not affect the intrinsic properties
related to the armchair edges along the length. Thus, \ce{C80H28} is
a suitable model for an armchair GNR of width parameter N=7, N being
the number of dimer lines across the ribbon width, according to the
standard notation \cite{son+06prl}. It corresponds to an effective
width of $\sim$ 7 \AA{}, defined as the distance between C atoms on
opposite edges, belonging to the same zigzag chain along $y$. Next,
we addressed the same GNF upon covalent edge functionalization with
organic groups. We started focusing on molecules containing a ketone
functional group, characterized by high electronegativity of its
carbonyl oxygen atoms, and in particular we considered the methyl
ketone group \mbox{-\ce{COCH3}} [inset in \ref{fig1}(b)], which is
known to be more stable than the simpler formaldehyde group -COH. We
simulated full functionalization of the \ce{C80H28} by replacing
each second H atom along the length with methyl ketone,
corresponding in this case to ten anchored groups per flake
[\ce{C80H18}\ce{(COCH3)10}], i.e. five per edge side, as shown in
\ref{fig1}(b). We computed $E_{G}$, EA and IP, and we compared these
quantities to those evaluated for \ce{C80H28} by calculating
$\Delta$EA ($\Delta$IP) as the difference between EA (IP) of the
functionalized and hydrogenated GNF, respectively. A positive shift
of both EA and IP is revealed, as shown in the second column of
\ref{table1}.

\begin{table*}
\renewcommand{\tabcolsep}{0.063cm}
\begin{tabular*}{\textwidth}{|l|c||c|c|c|c||c|c|}
\hline
\hline  & \includegraphics[scale=0.29]{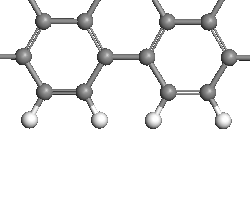} & \includegraphics[scale=0.29]{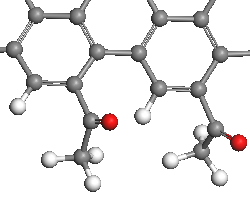} & \includegraphics[scale=0.29]{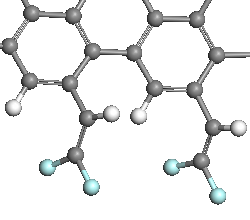} & \includegraphics[scale=0.29]{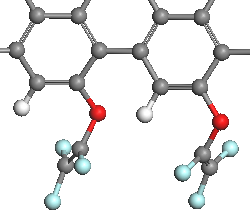} & \includegraphics[scale=0.29]{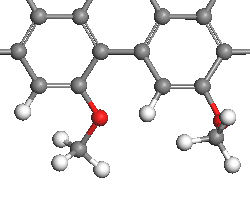} & \includegraphics[scale=0.29]{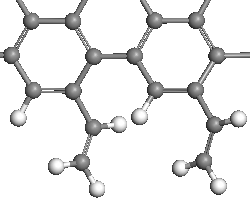}  & \includegraphics[scale=0.28]{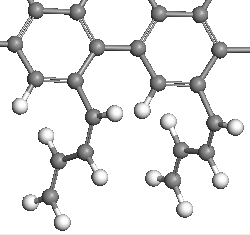} \\
\hline
& H & \mbox{-\ce{COCH3}} & \mbox{-\ce{CH=CF2}} & \mbox{-\ce{OCF=CF2}}
& \mbox{-\ce{OCH3}} & \mbox{-\ce{CH=CH2}} & \begin{small}\mbox{-\ce{(CH)3=CH2}}\end{small} \\
\hline
$E_{G}$ [eV] & 5.13 & 4.86 &4.65 & 4.74 &  4.75 &  4.76 & 4.67 \\
$\Delta$EA [eV] & 0.00 & 0.69 & 0.67 & 0.70 & -0.12 & 0.14 & 0.15 \\
$\Delta$IP [eV] & 0.00 & 0.42 & 0.19 & 0.31 & -0.50 & -0.22 & -0.31 \\
\hline \hline
\end{tabular*}
\caption{(Color online) Energy gap ($E_{G}$) and variations in EA
and IP ($\Delta$EA and $\Delta$IP), computed with respect to the
reference hydrogenated \ce{C80H28} (H), of GNFs fully functionalized
with different organic groups. Among the considered groups,
\mbox{-\ce{COCH3}}, \mbox{-\ce{CH=CF2}}, \mbox{-\ce{OCF=CF2}} and
\mbox{-\ce{OCH3}} give rise to a \textit{type II} offset, with a
positive (negative) sign of $\Delta$EA and $\Delta$IP for
electron-withdrawing (-donating) groups; \mbox{-\ce{CH=CH2}} and
\mbox{-\ce{(CH)3=CH2}} produce a \textit{type I} offset. }
\label{table1}
\end{table*}

In addition to the methyl ketone group, we considered the case of
other organic functionalizations. We focused on few prototypical
substituents, such as (i) a methoxy group (\mbox{-\ce{OCH3}}),
characterized by an O-bridge ester bond, which is known to be
electron-donating with respect to aromatic molecules
\cite{venk+07nl}; (ii) a \mbox{-\ce{CH=CF2}} group, which is
expected to be electron-withdrawing, due to the high
electronegativity of its F-termination; (iii) a \mbox{-\ce{OCF=CF2}}
group, which combines the presence of O-bridge ester bond and
halogen termination. Due to their characteristics, we expect group
(i) to provide a similar  effect compared to methyl ketone, i.e. to
rise  EA and IP with respect to the hydrogenated flake, and group
(ii) to do the opposite. In the case of \mbox{-\ce{OCF=CF2}} the
coexistence of both O-bridge bond and F-termination makes the
picture less predictable. We also consider \mbox{-\ce{CH=CH2}} and
\mbox{-\ce{(CH)3=CH2}} functionalizations which do not introduce any
foreign species in the system and preserve its C-$sp^2$ conjugation
through the alternation of single and double C-C bonds. For each
chosen group we addressed full functionalization, analogously to the
case of \mbox{-\ce{COCH3}} illustrated in \ref{fig1}(b), and we
computed $E_G$, EA and IP.

By inspecting the complete results in \ref{table1}, two main
features arise. The first is related to a decrease of the energy
gap, independently of the anchored group: this is ascribed to an
increase of the effective width of the flake, compared to the fully
hydrogenated case. The other important outcome concerns the sign of
$\Delta$EA and $\Delta$IP: we observe positive (negative) values of
both $\Delta$EA and $\Delta$IP, corresponding to a downshift
(upshift) of both HOMO and LUMO for groups with a predominant
electron-withdrawing (-donating) character, as well as for
\mbox{-\ce{OCF=CF2}}. On the contrary, \mbox{-\ce{CH=CH2}} and
\mbox{-\ce{(CH)3=CH2}} functionalizations produce  $\Delta$EA and
$\Delta$IP having opposite signs:
given the simple bond-alternating character of the functional group,
the major result is related to the extension of the wave function to
a larger effective width, which thus reduces $E_{G}$ in agreement
with results for one-dimensional GNRs~\cite{prez+10tobe}.

The analysis of Mulliken charge population shows that a charge redistribution at the edge occurs upon functionalization.
In particular Mulliken charges of functionalized C edge atoms become more positive, being these atoms bonded to more electronegative species than H.
Larger differences (> 0.2 $e^-$) are observed when functional groups are bonded to the flake through an O-bridge ester bond, i.e. \mbox{-\ce{OCH3}} and \mbox{-\ce{OCF=CF2}}.
On the other hand, moieties bonded through C atoms induce slighter Mulliken charge redistribution, of the order or less than 0.1 $e^-$.
It is important to highlight that these redistribution effects vanish within one aromatic ring.

\begin{table}[t!]
\renewcommand{\tabcolsep}{0.15cm}
\begin{tabular*}{.50\textwidth}{l|c|ccccc}
\hline
\hline  & H & 2 & 4 & 6 & 8 & 10 \\
\hline $E_{G}$ [eV] & 5.13  & 5.01 & 5.01 & 4.99 & 4.96 & 4.86 \\
$\Delta$EA [eV] & 0.00 & 0.21 & 0.29 & 0.37 & 0.47 & 0.69 \\
$\Delta$IP [eV] & 0.00  & 0.09 & 0.17 & 0.23 & 0.30 & 0.42 \\
\hline  \hline
\end{tabular*}
\caption{Energy gap ($E_{G}$) and variations in EA and IP ($\Delta$EA and $\Delta$IP) of the GNF functionalized with an increasing number of \mbox{-\ce{COCH3}} groups, up to 10 (full functionalization), with respect to hydrogenated \ce{C80H28} (H).}
\label{table2}
\end{table}

We also performed some tests on functionalized \ce{C80H28} at decreasing number of functional groups (see \ref{table2}). We focus here in \mbox{-\ce{COCH3}} as an
illustrative example but this behavior is independent of the
anchored group. We take the hydrogenated flake and, retaining the
symmetry of the system, we introduce increasing even numbers of
functional groups, up to ten, which corresponds to full
functionalization for the chosen \ce{C80H28} flake (see
\ref{fig1}(b)). We start from two anchored groups in positions 3 and
3', after the scheme in \ref{fig1}(a), which corresponds to a
\mbox{-\ce{COCH3}} substituent each fifth H atom. We then introduce
four and six groups in positions (2, 2', 4, 4') and  (1, 1', 3, 3',
5, 5') respectively.
In the case of eight bonded
groups, in order to maintain the symmetry of the graphene flake,
each H atom in positions (1, 1', 2, 2', 4, 4', 5, 5) is replaced by
a \mbox{-\ce{COCH3}}. The results, reported in \ref{table2}, show a
monotonic growth of both $\Delta$EA and $\Delta$IP with increasing
number of covalently bonded groups, with an almost linear dependence
for $\Delta$IP. At the same time, we notice the monotonic decrease
of $E_G$, consistently with the above mentioned picture related to
the increased effective width of the GNF.

We also discuss the dependence of the electronic properties of the
GNF, specifically related to EA and IP, by directly increasing its
width.
For this purpose, we performed a systematic
investigation on GNFs, characterized by fixed length (along $x$
direction in \ref{fig1}) and increasing width (along $y$
direction), fully functionalized with 16 \mbox{-\ce{COCH3}} groups \cite{lw-ratio}.
We studied eight GNFs ranging from width parameter N=4
(less than 4 \AA{} wide) to N=11 (about 12 \AA{} wide) and we computed EA and IP for each of them.
\begin{figure}%
\centering
\includegraphics[width=.48\textwidth]{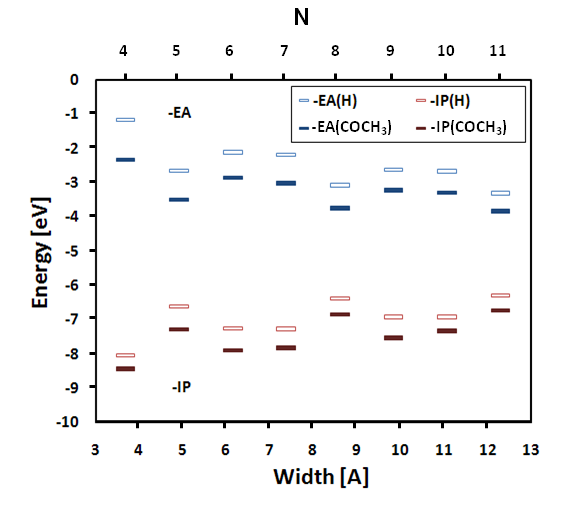}%
\caption{(Color online) Negative of electron affinity and ionization
potential (-EA and -IP, respectively) of hydrogenated (H) and
(\mbox{-\ce{COCH3}}) functionalized GNFs of increasing width. We
also indicate the parameter N, i.e. the number of C-C dimers along
the width.} \label{fig2}
\end{figure}
As shown in \ref{fig2}, the calculated values for EA and IP
have an oscillating behavior characterized by the \textit{modulo 3}
periodicity already encountered in infinite GNRs \cite{naka+96prb,baro+06nl,son+06prl},
with the smallest value of the energy gap pertaining to the N=$3p+2$
family ($p$ integer).
An approximate $\sim 1/w$ dependence (being $w$ the GNF width)  is
recognizable for the energy gap, consistently with well known
theoretical~\cite{son+06prl,yang+07prl} and
experimental~\cite{han+07prl,chen+07pe} results. The values of $\Delta$EA and
$\Delta$IP, represented in \ref{fig2} by the spacing between
the respective marks, show a faster decay, which may be compatible
with a local dipole mechanism driving the offset between
functionalized and hydrogenated flakes. For instance, for the $3p+2$
family, we get values of $\Delta$EA=0.85 eV and $\Delta$IP=0.67
eV for the smallest computed flake (N=5, $w$=4.9 \AA{}) and
$\Delta$EA=0.53 eV and $\Delta$IP=0.43 eV for the largest one
(N=11, $w$=12.3 \AA{}). From the results in \ref{fig2}, we
can extrapolate non negligible effects of covalent functionalization
also for nanometer-sized graphene flakes, which can be considered
realistic for current nanofabrication approaches.

Finally, we directly investigate the \textit{type II} nanojunction
obtained by connecting a fully hydrogenated GNF to a fully
functionalized one,  having the same width (see \ref{fig1}(c)). For
this purpose we adopted a flake of the same width of \ce{C80H28} but
increased global length, in order to ensure that each side of the
junction is long enough to confidently reproduce the \textit{bulk}
properties of the flake (i.e. same single-particle orbitals in the
energy range close to the energy gap). The resulting system
accommodates full functionalization of one half flake with eight
radicals, four per side (see \ref{fig1}(c)).
\begin{figure*}%
\centering
\includegraphics[width=\textwidth,clip]{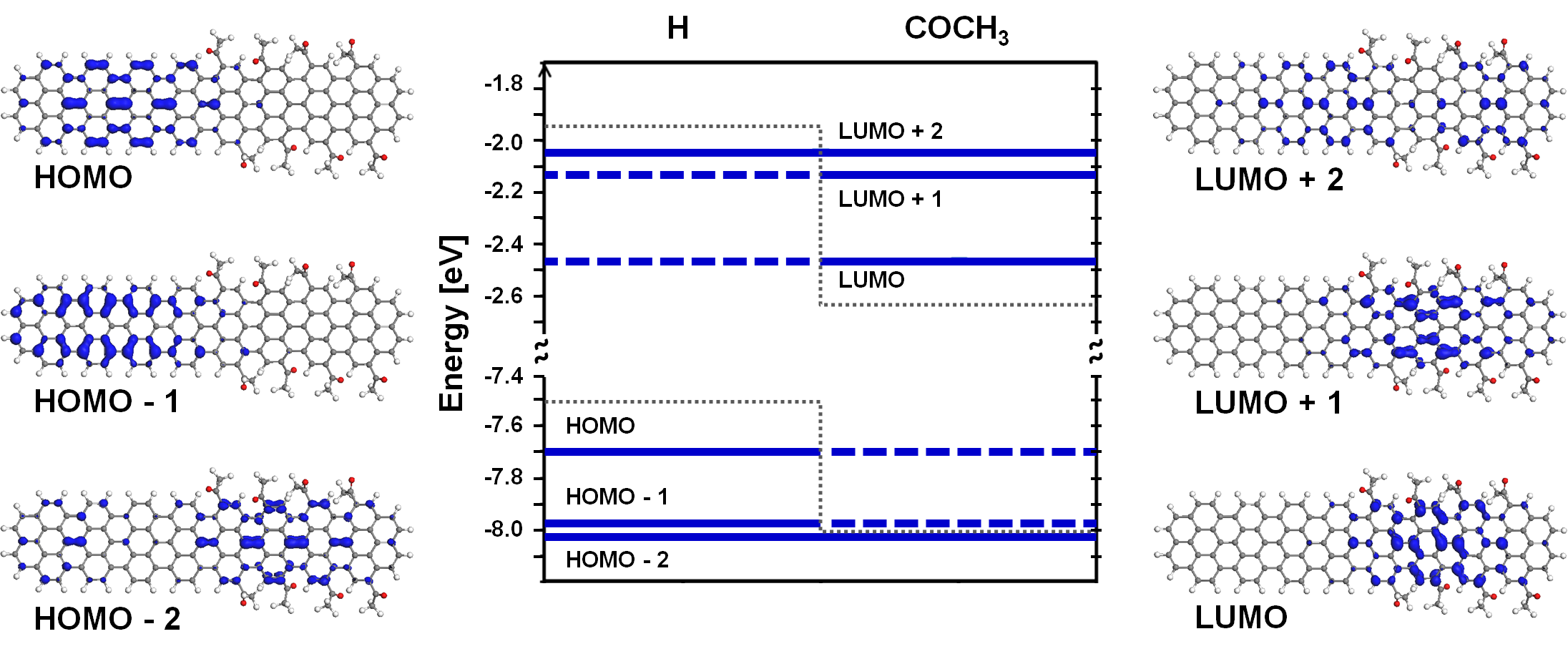}%
\caption{(Color online) Energy levels (center) and wave function
distributions  (sides) of the frontier states of the nanojunction
[\ref{fig1}(c)]. The dotted grey line in the energy level diagram
indicates the reference (from the homogeneous GNFs)  for HOMO and
LUMO of the hydrogenated (H) and functionalized (\mbox{-\ce{COCH3}})
sides of the GNF. The energy positions of the six frontier orbitals
represented at the side panels are marked by the heavier (blue)
lines: the solid intervals indicate finite wave function
probability, while dashed intervals indicate
vanishing wave function probability.}%
\label{fig3}
\end{figure*}
We first performed full geometrical optimization and then focused on
the ground state electronic properties of the junction. In
particular we are interested in the energy and wave function
localization of the frontier states, as summarized in \ref{fig3}.
The analysis of the wave function distributions clearly reflects the
\textit{type II} character of this junction. In fact we observe that
the top valence states, HOMO and HOMO-1, are localized on the
hydrogenated side of the junction, while the bottom conduction
states, LUMO and LUMO+1, are localized on the functionalized side.
On the other hand the next states HOMO-2 and LUMO+2 show already
resonant character. This feature can be explained in terms of the
absolute energy levels of the \textit{bulk} of the two sides, as
indicated in the graph of \ref{fig3}: we represent by the dotted
grey line the gap of the homogeneous systems, hydrogenated and
functionalized, evaluated from the HOMO and LUMO levels of the fully
hydrogenated [\ce{C116H40}] and functionalized
[\ce{C116H24}\ce{(COCH3)16}] GNFs. On top of this we plot the level
spectrum for the junction-GNF, highlighting the localization
character of the related state through blue solid (finite
probability) or dashed (zero probability) lines. We see that HOMO
and HOMO-1  lie in the forbidden gap for the functionalized side and
are consistently localized on the allowed hydrogenated half; the
opposite applies to LUMO and LUMO+1, which are thus localized on the
functionalized part. While the numbers for IP and EA in \ref{table2}
are taken from total energy calculations, therefore giving us good
precision, we remark that the energy level spectrum shown here is by
construction based on one-electron mean-field approximation and
cannot give us a level-by-level accuracy, in particular for the
unoccupied states. We see however that the overall picture is
totally consistent with the \textit{type II} alignment.

\section{Conclusions}
In conclusion, we have studied the effects of organic covalent
functionalization on  sub-nanometer sized graphene nanoflakes, by
varying the anchored groups, the number of
functionalizing groups and the flake width.
Our results indicate that edge functionalization can be designed to
tune the  electronic properties of graphene nanostructures and to
control band alignment in graphene nano-junctions, while preserving
stability and intrinsic electronic properties.
This opens the possibility to realize stable all-graphene electronic
and optoelectronic nanodevices based on \textit{type II} offset \cite{cocc+10tobe}.

\begin{acknowledgement}

The authors acknowledge CINECA for computational support and Stefano
Corni and Enrico Benassi for helpful discussions. This work was
partly supported by ``Fondazione Cassa di Risparmio di Modena'', by the
Italian Ministry of Foreign Affairs (Italy-USA bilateral project) and by the Italian Ministry of University and Research under FIRB grant ItalNanoNet.
MJC acknowledges support from FAPESP and CNPq (Brazil).

\end{acknowledgement}

\begin{mcitethebibliography}{56}
\providecommand*{\natexlab}[1]{#1}
\providecommand*{\mciteSetBstSublistMode}[1]{}
\providecommand*{\mciteSetBstMaxWidthForm}[2]{}
\providecommand*{\mciteBstWouldAddEndPuncttrue}
  {\def\EndOfBibitem{\unskip.}}
\providecommand*{\mciteBstWouldAddEndPunctfalse}
  {\let\EndOfBibitem\relax}
\providecommand*{\mciteSetBstMidEndSepPunct}[3]{}
\providecommand*{\mciteSetBstSublistLabelBeginEnd}[3]{}
\providecommand*{\EndOfBibitem}{}
\mciteSetBstSublistMode{f}
\mciteSetBstMaxWidthForm{subitem}{(\alph{mcitesubitemcount})}
\mciteSetBstSublistLabelBeginEnd{\mcitemaxwidthsubitemform\space}
{\relax}{\relax}

\bibitem[Novoselov et~al.(2004)Novoselov, Geim, Morozov, Jiang, Zhang, Dubonos,
  Grigorieva, and Firsov]{novo+04sci}
Novoselov,~K.~S.; Geim,~A.~K.; Morozov,~S.~V.; Jiang,~D.; Zhang,~Y.;
  Dubonos,~S.~V.; Grigorieva,~I.~V.; Firsov,~A.~A. Electric Field Effect in
  Atomically Thin Carbon Films. \emph{Science} \textbf{2004}, \emph{306},
  666--669\relax
\mciteBstWouldAddEndPuncttrue
\mciteSetBstMidEndSepPunct{\mcitedefaultmidpunct}
{\mcitedefaultendpunct}{\mcitedefaultseppunct}\relax
\EndOfBibitem
\bibitem[Soldano et~al.(2010)Soldano, Mahmood, and Dujardin]{sold+10car}
Soldano,~C.; Mahmood,~A.; Dujardin,~E. {Production, properties and potential of
  graphene}. \emph{Carbon} \textbf{2010}, \emph{48}, 2127--2150\relax
\mciteBstWouldAddEndPuncttrue
\mciteSetBstMidEndSepPunct{\mcitedefaultmidpunct}
{\mcitedefaultendpunct}{\mcitedefaultseppunct}\relax
\EndOfBibitem
\bibitem[Geim(2009)]{geim09sci}
Geim,~A. {Graphene: status and prospects}. \emph{Science} \textbf{2009},
  \emph{324}, 1530\relax
\mciteBstWouldAddEndPuncttrue
\mciteSetBstMidEndSepPunct{\mcitedefaultmidpunct}
{\mcitedefaultendpunct}{\mcitedefaultseppunct}\relax
\EndOfBibitem
\bibitem[Burghard et~al.(2009)Burghard, Klauk, and Kern]{burg+09am}
Burghard,~M.; Klauk,~H.; Kern,~K. {Carbon-Based Field-Effect Transistors for
  Nanoelectronics}. \emph{Adv.~Mater.~} \textbf{2009}, \emph{21},
  2586--2600\relax
\mciteBstWouldAddEndPuncttrue
\mciteSetBstMidEndSepPunct{\mcitedefaultmidpunct}
{\mcitedefaultendpunct}{\mcitedefaultseppunct}\relax
\EndOfBibitem
\bibitem[Castro~Neto et~al.(2009)Castro~Neto, Guinea, Peres, Novoselov, and
  Geim]{cast+09rmp}
Castro~Neto,~A.~H.; Guinea,~F.; Peres,~N. M.~R.; Novoselov,~K.~S.; Geim,~A.~K.
  The electronic properties of graphene. \emph{Rev.~Mod.~Phys.~} \textbf{2009},
  \emph{81}, 109--162\relax
\mciteBstWouldAddEndPuncttrue
\mciteSetBstMidEndSepPunct{\mcitedefaultmidpunct}
{\mcitedefaultendpunct}{\mcitedefaultseppunct}\relax
\EndOfBibitem
\bibitem[Han et~al.(2007)Han, Ozyilmaz, Zhang, and Kim]{han+07prl}
Han,~M.~Y.; Ozyilmaz,~B.; Zhang,~Y.; Kim,~P. Energy Band-Gap Engineering of
  Graphene Nanoribbons. \emph{Phys.~Rev.~Lett.~} \textbf{2007}, \emph{98},
  206805\relax
\mciteBstWouldAddEndPuncttrue
\mciteSetBstMidEndSepPunct{\mcitedefaultmidpunct}
{\mcitedefaultendpunct}{\mcitedefaultseppunct}\relax
\EndOfBibitem
\bibitem[Chen et~al.(2007)Chen, Lin, Rooks, and Avouris]{chen+07pe}
Chen,~Z.; Lin,~Y.-M.; Rooks,~M.~J.; Avouris,~P. Graphene nano-ribbon
  electronics. \emph{Physica E} \textbf{2007}, \emph{40}, 228 -- 232\relax
\mciteBstWouldAddEndPuncttrue
\mciteSetBstMidEndSepPunct{\mcitedefaultmidpunct}
{\mcitedefaultendpunct}{\mcitedefaultseppunct}\relax
\EndOfBibitem
\bibitem[Son et~al.(2006)Son, Cohen, and Louie]{son+06prl}
Son,~Y.-W.; Cohen,~M.~L.; Louie,~S.~G. Energy Gaps in Graphene Nanoribbons.
  \emph{Phys.~Rev.~Lett.~} \textbf{2006}, \emph{97}, 216803\relax
\mciteBstWouldAddEndPuncttrue
\mciteSetBstMidEndSepPunct{\mcitedefaultmidpunct}
{\mcitedefaultendpunct}{\mcitedefaultseppunct}\relax
\EndOfBibitem
\bibitem[Yang et~al.(2007)Yang, Park, Son, Cohen, and Louie]{yang+07prl}
Yang,~L.; Park,~C.-H.; Son,~Y.-W.; Cohen,~M.~L.; Louie,~S.~G. Quasiparticle
  Energies and Band Gaps in Graphene Nanoribbons. \emph{Phys.~Rev.~Lett.~}
  \textbf{2007}, \emph{99}, 186801\relax
\mciteBstWouldAddEndPuncttrue
\mciteSetBstMidEndSepPunct{\mcitedefaultmidpunct}
{\mcitedefaultendpunct}{\mcitedefaultseppunct}\relax
\EndOfBibitem
\bibitem[Prezzi et~al.(2008)Prezzi, Varsano, Ruini, Marini, and
  Molinari]{prez+08prb}
Prezzi,~D.; Varsano,~D.; Ruini,~A.; Marini,~A.; Molinari,~E. Optical properties
  of graphene nanoribbons: The role of many-body effects. \emph{Phys.~Rev.~B}
  \textbf{2008}, \emph{77}, 041404(R)\relax
\mciteBstWouldAddEndPuncttrue
\mciteSetBstMidEndSepPunct{\mcitedefaultmidpunct}
{\mcitedefaultendpunct}{\mcitedefaultseppunct}\relax
\EndOfBibitem
\bibitem[Prezzi et~al.(2007)Prezzi, Varsano, Ruini, Marini, and
  Molinari]{prez+07pssb}
Prezzi,~D.; Varsano,~D.; Ruini,~A.; Marini,~A.; Molinari,~E. Optical properties
  of one-dimensional graphene polymers: the case of polyphenanthrene.
  \emph{Phys.~Status~Solidi~B} \textbf{2007}, \emph{244}, 4124--4128\relax
\mciteBstWouldAddEndPuncttrue
\mciteSetBstMidEndSepPunct{\mcitedefaultmidpunct}
{\mcitedefaultendpunct}{\mcitedefaultseppunct}\relax
\EndOfBibitem
\bibitem[pre()]{prez+10tobe}
 Prezzi, D.; Varsano, D.; Ruini, A.; Molinari, E. submitted
  \textbf{2010}.\relax
\mciteBstWouldAddEndPunctfalse
\mciteSetBstMidEndSepPunct{\mcitedefaultmidpunct}
{}{\mcitedefaultseppunct}\relax
\EndOfBibitem
\bibitem[Yang et~al.(2007)Yang, Cohen, and Louie]{yang+07nl}
Yang,~L.; Cohen,~M.; Louie,~S. Excitonic Effects in the Optical Spectra of
  Graphene Nanoribbons. \emph{Nano~Lett.~} \textbf{2007}, \emph{7},
  3112--3115\relax
\mciteBstWouldAddEndPuncttrue
\mciteSetBstMidEndSepPunct{\mcitedefaultmidpunct}
{\mcitedefaultendpunct}{\mcitedefaultseppunct}\relax
\EndOfBibitem
\bibitem[Datta et~al.(2008)Datta, Strachan, Khamis, and Johnson]{datt+08nl}
Datta,~S.~S.; Strachan,~D.~R.; Khamis,~S.~M.; Johnson,~A. T.~C.
  Crystallographic Etching of Few-Layer Graphene. \emph{Nano~Lett.~}
  \textbf{2008}, \emph{8}, 1530--6984\relax
\mciteBstWouldAddEndPuncttrue
\mciteSetBstMidEndSepPunct{\mcitedefaultmidpunct}
{\mcitedefaultendpunct}{\mcitedefaultseppunct}\relax
\EndOfBibitem
\bibitem[Campos et~al.(2009)Campos, Manfrinato, Sanchez-Yamagishi, Kong, and
  Jarillo-Herrero]{camp+09nl}
Campos,~L.~C.; Manfrinato,~V.~R.; Sanchez-Yamagishi,~J.~D.; Kong,~J.;
  Jarillo-Herrero,~P. Anisotropic Etching and Nanoribbon Formation in
  Single-Layer Graphene. \emph{Nano~Lett.~} \textbf{2009}, \emph{9},
  1530--6984\relax
\mciteBstWouldAddEndPuncttrue
\mciteSetBstMidEndSepPunct{\mcitedefaultmidpunct}
{\mcitedefaultendpunct}{\mcitedefaultseppunct}\relax
\EndOfBibitem
\bibitem[Li et~al.(2008)Li, Wang, Zhang, Lee, and Dai]{li+08sci}
Li,~X.; Wang,~X.; Zhang,~L.; Lee,~S.; Dai,~H. {Chemically Derived, Ultrasmooth
  Graphene Nanoribbon Semiconductors}. \emph{Science} \textbf{2008},
  \emph{319}, 1229--1232\relax
\mciteBstWouldAddEndPuncttrue
\mciteSetBstMidEndSepPunct{\mcitedefaultmidpunct}
{\mcitedefaultendpunct}{\mcitedefaultseppunct}\relax
\EndOfBibitem
\bibitem[Wang and Dai(2010)]{wang-dai10natc}
Wang,~X.; Dai,~H. Etching and narrowing of graphene from the edges.
  \emph{Nature Chem.} \textbf{2010}, \emph{2}, 661--665\relax
\mciteBstWouldAddEndPuncttrue
\mciteSetBstMidEndSepPunct{\mcitedefaultmidpunct}
{\mcitedefaultendpunct}{\mcitedefaultseppunct}\relax
\EndOfBibitem
\bibitem[Jia et~al.(2009)Jia, Hofmann, Meunier, Sumpter, Campos-Delgado,
  Romo-Herrera, Son, Hsieh, Reina, Kong, Terrones, and Dresselhaus]{jia+09sci}
Jia,~X.; Hofmann,~M.; Meunier,~V.; Sumpter,~B.; Campos-Delgado,~J.;
  Romo-Herrera,~J.; Son,~H.; Hsieh,~Y.-P.; Reina,~A.; Kong,~J.; Terrones,~M.;
  Dresselhaus,~M. Controlled Formation of Sharp Zigzag and Armchair Edges in
  Graphitic Nanoribbons. \emph{Science} \textbf{2009}, \emph{323}, 1701\relax
\mciteBstWouldAddEndPuncttrue
\mciteSetBstMidEndSepPunct{\mcitedefaultmidpunct}
{\mcitedefaultendpunct}{\mcitedefaultseppunct}\relax
\EndOfBibitem
\bibitem[Kosynkin et~al.(2009)Kosynkin, Higginbotham, Sinitskii, Lomeda,
  Dimiev, Price, and Tour]{kosy+09nat}
Kosynkin,~D.~V.; Higginbotham,~A.~L.; Sinitskii,~A.; Lomeda,~J.~R.; Dimiev,~A.;
  Price,~B.~K.; Tour,~J.~M. Longitudinal unzipping of carbon nanotubes to form
  graphene nanoribbons. \emph{Nature~(London)~} \textbf{2009}, \emph{458},
  872\relax
\mciteBstWouldAddEndPuncttrue
\mciteSetBstMidEndSepPunct{\mcitedefaultmidpunct}
{\mcitedefaultendpunct}{\mcitedefaultseppunct}\relax
\EndOfBibitem
\bibitem[Jiao et~al.(2009)Jiao, Zhang, Wang, Diankov, and Dai]{jiao+09nat}
Jiao,~L.; Zhang,~L.; Wang,~X.; Diankov,~G.; Dai,~H. Narrow graphene nanoribbons
  from carbon nanotubes. \emph{Nature~(London)~} \textbf{2009}, \emph{458},
  877\relax
\mciteBstWouldAddEndPuncttrue
\mciteSetBstMidEndSepPunct{\mcitedefaultmidpunct}
{\mcitedefaultendpunct}{\mcitedefaultseppunct}\relax
\EndOfBibitem
\bibitem[Cano-M\'arquez et~al.(2009)Cano-M\'arquez, Rodr\'\i{}guez-Mac\'\i{}as,
  Campos-Delgado, Espinosa-Gonz\'alez, Trist\'an-L\'opez,
  Ram\'\i{}rez-Gonz\'alez, Cullen, Smith, Terrones, , and
  Vega-Cant\'u]{cano+09nl}
Cano-M\'arquez,~A.~G.; Rodr\'\i{}guez-Mac\'\i{}as,~F.~J.; Campos-Delgado,~J.;
  Espinosa-Gonz\'alez,~C.~G.; Trist\'an-L\'opez,~F.;
  Ram\'\i{}rez-Gonz\'alez,~D.; Cullen,~D.~A.; Smith,~D.~J.; Terrones,~M.; ;
  Vega-Cant\'u,~Y.~I. Ex-MWNTs: Graphene Sheets and Ribbons Produced by Lithium
  Intercalation and Exfoliation of Carbon Nanotubes. \emph{Nano~Lett.~}
  \textbf{2009}, \emph{9}, 1527\relax
\mciteBstWouldAddEndPuncttrue
\mciteSetBstMidEndSepPunct{\mcitedefaultmidpunct}
{\mcitedefaultendpunct}{\mcitedefaultseppunct}\relax
\EndOfBibitem
\bibitem[Yang et~al.(2008)Yang, Dou, Rouhanipour, Zhi, Rader, and
  Muellen]{yang+08jacs}
Yang,~X.; Dou,~X.; Rouhanipour,~A.; Zhi,~L.; Rader,~H.~J.; Muellen,~K.
  Two-Dimensional Graphene Nanoribbons. \emph{J.~Am.~Chem.~Soc.~}
  \textbf{2008}, \emph{130}, 4216\relax
\mciteBstWouldAddEndPuncttrue
\mciteSetBstMidEndSepPunct{\mcitedefaultmidpunct}
{\mcitedefaultendpunct}{\mcitedefaultseppunct}\relax
\EndOfBibitem
\bibitem[Cai et~al.(2010)Cai, Ruffieux, Jaafar, Bieri, Braun, Blankenburg,
  Muoth, Saleh, Feng, Muellen, and Fasel]{cai+10nat}
Cai,~J.; Ruffieux,~P.; Jaafar,~R.; Bieri,~M.; Braun,~T.; Blankenburg,~S.;
  Muoth,~A.~P.,~Matthiasand~Seitsonen; Saleh,~M.; Feng,~X.; Muellen,~K.;
  Fasel,~R. Atomically precise bottom-up fabrication of graphene nanoribbons.
  \emph{Nature~(London)~} \textbf{2010}, \emph{466}, 470--473\relax
\mciteBstWouldAddEndPuncttrue
\mciteSetBstMidEndSepPunct{\mcitedefaultmidpunct}
{\mcitedefaultendpunct}{\mcitedefaultseppunct}\relax
\EndOfBibitem
\bibitem[Wang et~al.(2008)Wang, Ouyang, Li, Wang, Guo, and Dai]{wang+08prl}
Wang,~X.; Ouyang,~Y.; Li,~X.; Wang,~H.; Guo,~J.; Dai,~H. Room-Temperature
  All-Semiconducting Sub-10-nm Graphene Nanoribbon Field-Effect Transistors.
  \emph{Phys.~Rev.~Lett.~} \textbf{2008}, \emph{100}, 206803\relax
\mciteBstWouldAddEndPuncttrue
\mciteSetBstMidEndSepPunct{\mcitedefaultmidpunct}
{\mcitedefaultendpunct}{\mcitedefaultseppunct}\relax
\EndOfBibitem
\bibitem[Xia et~al.(2009)Xia, Mueller, Lin, Valdes-Garcia, and
  Avouris]{xia+09natn}
Xia,~F.; Mueller,~T.; Lin,~Y.; Valdes-Garcia,~A.; Avouris,~P. {Ultrafast
  graphene photodetector}. \emph{Nature Nanotech.} \textbf{2009}, \emph{4},
  839--843\relax
\mciteBstWouldAddEndPuncttrue
\mciteSetBstMidEndSepPunct{\mcitedefaultmidpunct}
{\mcitedefaultendpunct}{\mcitedefaultseppunct}\relax
\EndOfBibitem
\bibitem[Ponomarenko et~al.(2008)Ponomarenko, Schedin, Katsnelson, Yang, Hill,
  Novoselov, and Geim]{pono+08sci}
Ponomarenko,~L.; Schedin,~F.; Katsnelson,~M.; Yang,~R.; Hill,~E.;
  Novoselov,~K.; Geim,~A. {Chaotic Dirac billiard in graphene quantum dots}.
  \emph{Science} \textbf{2008}, \emph{320}, 356\relax
\mciteBstWouldAddEndPuncttrue
\mciteSetBstMidEndSepPunct{\mcitedefaultmidpunct}
{\mcitedefaultendpunct}{\mcitedefaultseppunct}\relax
\EndOfBibitem
\bibitem[Loh et~al.(2010)Loh, Bao, Ang, and Yang]{loh+10jmc}
Loh,~K.; Bao,~Q.; Ang,~P.; Yang,~J. {The chemistry of graphene}.
  \emph{J.~Mater.~Chem.~} \textbf{2010}, \emph{20}, 2277--2289\relax
\mciteBstWouldAddEndPuncttrue
\mciteSetBstMidEndSepPunct{\mcitedefaultmidpunct}
{\mcitedefaultendpunct}{\mcitedefaultseppunct}\relax
\EndOfBibitem
\bibitem[Farmer et~al.(2009)Farmer, Golizadeh-Mojarad, Perebeinos, Lin,
  Tulevski, Tsang, and Avouris]{farm+09nl}
Farmer,~D.~B.; Golizadeh-Mojarad,~R.; Perebeinos,~V.; Lin,~Y.-M.;
  Tulevski,~G.~S.; Tsang,~J.~C.; Avouris,~P. Chemical doping and electron-hole
  conduction asymmetry in graphene devices. \emph{Nano~Lett.~} \textbf{2009},
  \emph{9}, 388--392\relax
\mciteBstWouldAddEndPuncttrue
\mciteSetBstMidEndSepPunct{\mcitedefaultmidpunct}
{\mcitedefaultendpunct}{\mcitedefaultseppunct}\relax
\EndOfBibitem
\bibitem[Farmer et~al.(2009)Farmer, Lin, Afzali-Ardakani, and
  Avouris]{farm+09apl}
Farmer,~D.~B.; Lin,~Y.-M.; Afzali-Ardakani,~A.; Avouris,~P. Behavior of a
  chemically doped graphene junction. \emph{Appl.~Phys.~Lett.~} \textbf{2009},
  \emph{94}, 213106\relax
\mciteBstWouldAddEndPuncttrue
\mciteSetBstMidEndSepPunct{\mcitedefaultmidpunct}
{\mcitedefaultendpunct}{\mcitedefaultseppunct}\relax
\EndOfBibitem
\bibitem[Lohmann et~al.(2009)Lohmann, von Klitzing, and Smet]{lohm+09nl}
Lohmann,~T.; von Klitzing,~K.; Smet,~J. {Four-Terminal Magneto-Transport in
  Graphene pn Junctions Created by Spatially Selective Doping}.
  \emph{Nano~Lett.~} \textbf{2009}, \emph{9}, 1973--1979\relax
\mciteBstWouldAddEndPuncttrue
\mciteSetBstMidEndSepPunct{\mcitedefaultmidpunct}
{\mcitedefaultendpunct}{\mcitedefaultseppunct}\relax
\EndOfBibitem
\bibitem[Lopez-Bezanilla et~al.(2009)Lopez-Bezanilla, Triozon, and
  Roche]{lope+09nl}
Lopez-Bezanilla,~A.; Triozon,~F.; Roche,~S. Chemical functionalization effects
  on armchair graphene nanoribbon transport. \emph{Nano~Lett.~} \textbf{2009},
  \emph{9}, 2537\relax
\mciteBstWouldAddEndPuncttrue
\mciteSetBstMidEndSepPunct{\mcitedefaultmidpunct}
{\mcitedefaultendpunct}{\mcitedefaultseppunct}\relax
\EndOfBibitem
\bibitem[Bekyarova et~al.(2009)Bekyarova, Itkis, Ramesh, Berger, Sprinkle,
  de~Heer, and Haddon]{beky+09jacs}
Bekyarova,~E.; Itkis,~M.; Ramesh,~P.; Berger,~C.; Sprinkle,~M.; de~Heer,~W.;
  Haddon,~R. {Chemical Modification of Epitaxial Graphene: Spontaneous Grafting
  of Aryl Groups}. \emph{J.~Am.~Chem.~Soc.~} \textbf{2009}, \emph{131},
  1336--1337\relax
\mciteBstWouldAddEndPuncttrue
\mciteSetBstMidEndSepPunct{\mcitedefaultmidpunct}
{\mcitedefaultendpunct}{\mcitedefaultseppunct}\relax
\EndOfBibitem
\bibitem[Sinitskii et~al.(2010)Sinitskii, Dimiev, Corley, Fursina, Kosynkin,
  and Tour]{sini+10nano}
Sinitskii,~A.; Dimiev,~A.; Corley,~D.; Fursina,~A.; Kosynkin,~D.; Tour,~J.
  {Kinetics of Diazonium Functionalization of Chemically Converted Graphene
  Nanoribbons}. \emph{ACS~Nano} \textbf{2010}, \emph{4}, 1949--1954\relax
\mciteBstWouldAddEndPuncttrue
\mciteSetBstMidEndSepPunct{\mcitedefaultmidpunct}
{\mcitedefaultendpunct}{\mcitedefaultseppunct}\relax
\EndOfBibitem
\bibitem[Chen et~al.(2007)Chen, Chen, Qi, Gao, and Wee]{chen+07jacs}
Chen,~W.; Chen,~S.; Qi,~D.~C.; Gao,~X.~Y.; Wee,~A. T.~S. Surface Transfer
  p-Type Doping of Epitaxial Graphene. \emph{J.~Am.~Chem.~Soc.~} \textbf{2007},
  \emph{129}, 10418--10422\relax
\mciteBstWouldAddEndPuncttrue
\mciteSetBstMidEndSepPunct{\mcitedefaultmidpunct}
{\mcitedefaultendpunct}{\mcitedefaultseppunct}\relax
\EndOfBibitem
\bibitem[Lu et~al.(2009)Lu, Chen, Feng, and He]{lu+09jpcb}
Lu,~Y.~H.; Chen,~W.; Feng,~Y.~P.; He,~P.~M. Tuning the Electronic Structure of
  Graphene by an Organic Molecule. \emph{J.~Phys.~Chem.~B} \textbf{2009},
  \emph{113}, 2--5\relax
\mciteBstWouldAddEndPuncttrue
\mciteSetBstMidEndSepPunct{\mcitedefaultmidpunct}
{\mcitedefaultendpunct}{\mcitedefaultseppunct}\relax
\EndOfBibitem
\bibitem[Zhang et~al.(2010)Zhang, Zhou, Xie, Zeng, Zhang, and Peng]{zhan+10nt}
Zhang,~Y.-H.; Zhou,~K.-G.; Xie,~K.-F.; Zeng,~J.; Zhang,~H.-L.; Peng,~Y. Tuning
  the electronic structure and transport properties of graphene by noncovalent
  functionalization: effects of organic donor, acceptor and metal atoms.
  \emph{Nanotechnol.~} \textbf{2010}, \emph{21}, 065201\relax
\mciteBstWouldAddEndPuncttrue
\mciteSetBstMidEndSepPunct{\mcitedefaultmidpunct}
{\mcitedefaultendpunct}{\mcitedefaultseppunct}\relax
\EndOfBibitem
\bibitem[Sun et~al.(2010)Sun, Lu, Chen, Feng, and Wee]{sun+10prb}
Sun,~J.~T.; Lu,~Y.~H.; Chen,~W.; Feng,~Y.~P.; Wee,~A. T.~S. Linear tuning of
  charge carriers in graphene by organic molecules and charge-transfer
  complexes. \emph{Phys.~Rev.~B} \textbf{2010}, \emph{81}, 155403\relax
\mciteBstWouldAddEndPuncttrue
\mciteSetBstMidEndSepPunct{\mcitedefaultmidpunct}
{\mcitedefaultendpunct}{\mcitedefaultseppunct}\relax
\EndOfBibitem
\bibitem[Qian et~al.(2008)Qian, Negri, Wang, and Wang]{qian+08jacs}
Qian,~H.; Negri,~F.; Wang,~C.; Wang,~Z. Fully Conjugated Tri(perylene
  bisimides): An Approach to the Construction of n-Type Graphene Nanoribbons.
  \emph{J.~Am.~Chem.~Soc.~} \textbf{2008}, \emph{130}, 17970\relax
\mciteBstWouldAddEndPuncttrue
\mciteSetBstMidEndSepPunct{\mcitedefaultmidpunct}
{\mcitedefaultendpunct}{\mcitedefaultseppunct}\relax
\EndOfBibitem
\bibitem[Wang et~al.(2009)Wang, Li, Zhang, Yoon, Weber, Wang, Guo, and
  Dai]{wang+09sci}
Wang,~X.; Li,~X.; Zhang,~L.; Yoon,~Y.; Weber,~P.~K.; Wang,~H.; Guo,~J.; Dai,~H.
  N-Doping of Graphene Through Electrothermal Reactions with Ammonia.
  \emph{Science} \textbf{2009}, \emph{324}, 768--771\relax
\mciteBstWouldAddEndPuncttrue
\mciteSetBstMidEndSepPunct{\mcitedefaultmidpunct}
{\mcitedefaultendpunct}{\mcitedefaultseppunct}\relax
\EndOfBibitem
\bibitem[Sevin{\c{c}}li et~al.(2008)Sevin{\c{c}}li, Topsakal, and
  Ciraci]{sevi+08prb}
Sevin{\c{c}}li,~H.; Topsakal,~M.; Ciraci,~S. {Superlattice structures of
  graphene-based armchair nanoribbons}. \emph{Phys.~Rev.~B} \textbf{2008},
  \emph{78}, 245402\relax
\mciteBstWouldAddEndPuncttrue
\mciteSetBstMidEndSepPunct{\mcitedefaultmidpunct}
{\mcitedefaultendpunct}{\mcitedefaultseppunct}\relax
\EndOfBibitem
\bibitem[Dewar et~al.(1985)Dewar, Zoebish, Healy, and Stewart]{dewa+85jacs}
Dewar,~M. J.~S.; Zoebish,~E.~G.; Healy,~E.~F.; Stewart,~J. J.~P. A new general
  purpose quantum mechanical molecular model. \emph{J.~Am.~Chem.~Soc.~}
  \textbf{1985}, \emph{107}, 3902--3909\relax
\mciteBstWouldAddEndPuncttrue
\mciteSetBstMidEndSepPunct{\mcitedefaultmidpunct}
{\mcitedefaultendpunct}{\mcitedefaultseppunct}\relax
\EndOfBibitem
\bibitem[AM1()]{AM1-note}
 We used AM1 as implemented in the VAMP package, included in Accelrys Materials
  Studio software, version 5.0. See also:
  \url{http://accelrys.com/products/materials-studio}\relax
\mciteBstWouldAddEndPuncttrue
\mciteSetBstMidEndSepPunct{\mcitedefaultmidpunct}
{\mcitedefaultendpunct}{\mcitedefaultseppunct}\relax
\EndOfBibitem
\bibitem[Roothaan(1951)]{root51rmp}
Roothaan,~C. C.~J. New developments in molecular orbital theory.
  \emph{Rev.~Mod.~Phys.~} \textbf{1951}, \emph{23}, 69\relax
\mciteBstWouldAddEndPuncttrue
\mciteSetBstMidEndSepPunct{\mcitedefaultmidpunct}
{\mcitedefaultendpunct}{\mcitedefaultseppunct}\relax
\EndOfBibitem
\bibitem[Caldas et~al.(2001)Caldas, Pettenati, Goldoni, and
  Molinari]{cald+01apl}
Caldas,~M.~J.; Pettenati,~E.; Goldoni,~G.; Molinari,~E. Tailoring of light
  emission properties of functionalized oligothiophenes.
  \emph{Appl.~Phys.~Lett.~} \textbf{2001}, \emph{79}, 2505--2507\relax
\mciteBstWouldAddEndPuncttrue
\mciteSetBstMidEndSepPunct{\mcitedefaultmidpunct}
{\mcitedefaultendpunct}{\mcitedefaultseppunct}\relax
\EndOfBibitem
\bibitem[D{\'a}vila and Caldas(2002)]{davi-cald02jcc}
D{\'a}vila,~L. Y.~A.; Caldas,~M.~J. Applicability of MNDO techniques AM1 and
  PM3 to Ring-Structured Polymers. \emph{J.~Comput.~Chem.~} \textbf{2002},
  \emph{23}, 1135\relax
\mciteBstWouldAddEndPuncttrue
\mciteSetBstMidEndSepPunct{\mcitedefaultmidpunct}
{\mcitedefaultendpunct}{\mcitedefaultseppunct}\relax
\EndOfBibitem
\bibitem[Lair et~al.(2008)Lair, Herndon, and Murr]{lair+08car}
Lair,~S.; Herndon,~W.; Murr,~L. {Stability comparison of simulated
  double-walled carbon nanotube structures}. \emph{Carbon} \textbf{2008},
  \emph{46}, 2083--2095\relax
\mciteBstWouldAddEndPuncttrue
\mciteSetBstMidEndSepPunct{\mcitedefaultmidpunct}
{\mcitedefaultendpunct}{\mcitedefaultseppunct}\relax
\EndOfBibitem
\bibitem[Tachikawa(2008)]{tach08jpcc}
Tachikawa,~H. A direct molecular orbital-molecular dynamics study on the
  diffusion of the Li ion on a fluorinated graphene surface.
  \emph{J.~Phys.~Chem.~C} \textbf{2008}, \emph{112}, 10193--10199\relax
\mciteBstWouldAddEndPuncttrue
\mciteSetBstMidEndSepPunct{\mcitedefaultmidpunct}
{\mcitedefaultendpunct}{\mcitedefaultseppunct}\relax
\EndOfBibitem
\bibitem[Hantal et~al.(2009)Hantal, Picaud, Collignon, Hoang, Rayez, and
  Rayez]{hant+09ms}
Hantal,~G.; Picaud,~S.; Collignon,~B.; Hoang,~P. N.~M.; Rayez,~M.~T.;
  Rayez,~J.~C. A new semi-empirical model for the oxidation of PAHs physisorbed
  on soot. I. Application to the reaction $C_{6}H_{6}$+OH. \emph{Mol. Simul.}
  \textbf{2009}, \emph{35}, 1130\relax
\mciteBstWouldAddEndPuncttrue
\mciteSetBstMidEndSepPunct{\mcitedefaultmidpunct}
{\mcitedefaultendpunct}{\mcitedefaultseppunct}\relax
\EndOfBibitem
\bibitem[Abe et~al.(2010)Abe, Nagoya, Watari, and Tachikawa]{abe+10jjap}
Abe,~S.; Nagoya,~Y.; Watari,~F.; Tachikawa,~H. Interaction of Water Molecules
  with Graphene: A Density Functional Theory and Molecular Dynamics Study.
  \emph{Jpn.~J.~Appl.~Phys.~} \textbf{2010}, \emph{49}, 01AH07\relax
\mciteBstWouldAddEndPuncttrue
\mciteSetBstMidEndSepPunct{\mcitedefaultmidpunct}
{\mcitedefaultendpunct}{\mcitedefaultseppunct}\relax
\EndOfBibitem
\bibitem[Wetmore et~al.(2000)Wetmore, Boyd, and Eriksson]{wetm+00cpl}
Wetmore,~S.~D.; Boyd,~R.~J.; Eriksson,~L.~A. {Electron affinities and
  ionization potentials of nucleotide bases}. \emph{Chem.~Phys.~Lett.~}
  \textbf{2000}, \emph{322}, 129 -- 135\relax
\mciteBstWouldAddEndPuncttrue
\mciteSetBstMidEndSepPunct{\mcitedefaultmidpunct}
{\mcitedefaultendpunct}{\mcitedefaultseppunct}\relax
\EndOfBibitem
\bibitem[Venkataraman et~al.(2007)Venkataraman, Park, Whalley, Nuckolls,
  Hybertsen, and Steigerwald]{venk+07nl}
Venkataraman,~L.; Park,~Y.; Whalley,~A.; Nuckolls,~C.; Hybertsen,~M.;
  Steigerwald,~M. {Electronics and chemistry: varying single-molecule junction
  conductance using chemical substituents}. \emph{Nano~Lett.~} \textbf{2007},
  \emph{7}, 502--506\relax
\mciteBstWouldAddEndPuncttrue
\mciteSetBstMidEndSepPunct{\mcitedefaultmidpunct}
{\mcitedefaultendpunct}{\mcitedefaultseppunct}\relax
\EndOfBibitem
\bibitem[lw-()]{lw-ratio}
 To keep an appropriate length/width ratio for all structures, we considered a
  class of flakes having the same length of the nanojunction
  [\ref{fig1}(c)].\relax
\mciteBstWouldAddEndPunctfalse
\mciteSetBstMidEndSepPunct{\mcitedefaultmidpunct}
{}{\mcitedefaultseppunct}\relax
\EndOfBibitem
\bibitem[Nakada et~al.(1996)Nakada, Fujita, Dresselhaus, and
  Dresselhaus]{naka+96prb}
Nakada,~K.; Fujita,~M.; Dresselhaus,~G.; Dresselhaus,~M.~S. Edge state in
  graphene ribbons: Nanometer size effect and edge shape dependence.
  \emph{Phys.~Rev.~B} \textbf{1996}, \emph{54}, 17954--17961\relax
\mciteBstWouldAddEndPuncttrue
\mciteSetBstMidEndSepPunct{\mcitedefaultmidpunct}
{\mcitedefaultendpunct}{\mcitedefaultseppunct}\relax
\EndOfBibitem
\bibitem[Barone et~al.(2006)Barone, Hod, and Scuseria]{baro+06nl}
Barone,~V.; Hod,~O.; Scuseria,~G.~E. Electronic Structure and Stability of
  Semiconducting Graphene Nanoribbons. \emph{Nano~Lett.~} \textbf{2006},
  \emph{6}, 2748--2754\relax
\mciteBstWouldAddEndPuncttrue
\mciteSetBstMidEndSepPunct{\mcitedefaultmidpunct}
{\mcitedefaultendpunct}{\mcitedefaultseppunct}\relax
\EndOfBibitem
\bibitem[coc()]{cocc+10tobe}
 Cocchi, C.; Prezzi, D.; Ruini, A.; Caldas, M. J; Molinari, E.
  \textit{preprint} \textbf{2010}.\relax
\mciteBstWouldAddEndPunctfalse
\mciteSetBstMidEndSepPunct{\mcitedefaultmidpunct}
{}{\mcitedefaultseppunct}\relax
\EndOfBibitem
\end{mcitethebibliography}
\providecommand*{\mcitethebibliography}{\thebibliography}
\csname @ifundefined\endcsname{endmcitethebibliography}
{\let\endmcitethebibliography\endthebibliography}{}


\end{document}